\documentclass{article}
\input epsf.sty
\input colordvi.sty

\usepackage[]{graphicx}

\newcommand{\half}{\frac{1}{2}}
\newcommand{\bright}{\begin{flushright}}
\newcommand{\eright}{\end{flushright}}
\newcommand{\bminip}{\begin{minipage}}
\newcommand{\eminip}{\end{minipage}}
\newcommand{\bcent}{\begin{center}}
\newcommand{\ecent}{\end{center}}
\newcommand{\beq}{\begin{equation}}
\newcommand{\eeq}{\end{equation}}
\newcommand{\beqa}{\begin{eqnarray}}
\newcommand{\eeqa}{\end{eqnarray}}
\newcommand{\barr}{\begin{array}}
\newcommand{\earr}{\end{array}}
\renewcommand{\thefootnote}{\fnsymbol{footnote}}

\newcommand{\reflef}{(\ref}
\newcommand{\MP}{M_{\rm P}}

\newcommand{\Lmd}{\Lambda}

\newcommand{\psibar}{\overline{\psi}}

\begin{document}
%\mbox{}\\[-4.2em]
%\bright
%%\mbox{}\\[-2.6em]
%/texdoc/ipmu/ksh1c.tex \today
%\eright
\baselineskip=0.6cm
\mbox{}\\[-5.5em]
\bcent
{\Large\bf How successful can the scalar-tensor theory be in
understanding the accelerating universe?\footnote{Based on the talk
delivered on the occasion of IPMU international conference Dark Energy:
Lighting up the darkness!  Kashiwa, Japan 22-26 June 2009.}
}\\
Yasunori Fujii\\
Advanced Research Institute for Science and Engineering, Waseda University,\\[-.5em]
Okubo, Tokyo, 169-8555 Japan
\ecent
\mbox{}\\[-3.1em]
\bcent
\bminip{14cm}
\bcent
{\large\bf Abstract}
%\mbox{}\\[-6.9em]
\ecent
\mbox{}\\[-1.9em]
The accelerating universe is closely related to today's version of the
cosmological constant problem; fine-tuning and coincidence problems.
We show how successfully the scalar-tensor theory, a rather rigid
theoretical idea, provides us with a simple and natural way to understand why 
today's observed cosmological constant is small only because we are old
cosmologically, without fine-tuning theoretical parameters extremely.  
\eminip
\ecent

\renewcommand{\thefootnote}{\arabic{footnote}}
\setcounter{footnote}{0}
\mbox{}\\[-3.5em]

\section{Introduction} 

Nearly a decade ago, the universe was discovered to be accelerating,
driven by the cosmological constant represented by \cite{accu}
%%%%%%%%%%
\beq
\Omega_\Lmd =\Lmd_{\rm obs}/\rho_{\rm cr}\approx 0.7,\quad\mbox{with}\quad\rho_{\rm cr} =3H_0^2,
\label{ipmu_c1}
\eeq
where $H_0$ is today's value of the Hubble parameter.  Note also that we
are using the reduced Planckian units  with $c=\hbar =\MP (=(8\pi
G)^{-1/2})=1$.  We then expect $H_0 \sim t_0^{-1}$, hence
%%%%%%%%%%
\beq
\Lmd_{\rm obs} \sim t_0^{-2}, 
\label{ipmu_c2}
\eeq
where the present age of the universe is given by $t_0 =1.37 \times
10^{10}{\rm y}\approx 10^{60.2}$ in units of the Planck time.  The
result $\Lmd_{\rm obs} \sim 10^{-120}$ poses a fine-tuning problem in 
today's version of the cosmological constant problem: Is our theory good
enough to the accuracy of 120 orders of magnitude?  In addition, the relation
 \reflef{ipmu_c2}), to be called $\Lmd$-$t_0$ Correlation, raises an 
 even more serious question: Can we understand the numerical similarity
 between both sides beyond a mere coincidence?  Is this supported by any
 fundamental theory?  We will provide an affirmative reply in terms of
 the scalar-tensor theory invented first by Jordan in 1955
 \cite{jordan, cmb}.

\section{Scalar-tensor theory}

The basic Lagrangian in what is called the Jordan (conformal) frame is \cite{jordan}
%%%%%%%%%%
\beq
{\cal L}= \sqrt{-g} \hspace{-.0em}\left( \half \xi \phi^2 R -\epsilon \half g^{\mu\nu}\partial_\mu\phi \partial_\nu\phi -\Lmd \ +L_{\rm matter} \right), 
\label{ipmu_8}
\eeq
where $\phi$ is the scalar field, assumed to show itself as dark energy.
We use two parameters $\epsilon$ and $\xi$ related to the better known
symbol $\omega$ in the way of $\epsilon =\pm 1 ={\rm Sgn}(\omega)$ and $
\xi =1/(4|\omega| ) >0$.  The first term is a well-known nonminimal
coupling term with the effective gravitational ``constant,'' $
G_{\rm eff}(x)=( 8\pi\xi\phi^2)^{-1}$.  We included $-\Lmd$.   On the matter
Lagrangian $L_{\rm matter}$ we have a special comment. In 1961, Brans and Dicke
came up with an additional assumption that $\phi$ never enters $L_{\rm
matter}$, because they could save the idea of Weak Equivalence Principle
(WEP) only in this way \cite{BD}.  Since then the word ``Brans-Dicke
theory'' has been used widely for the entire theory, though we prefer to use
another name, the ``Brans-Dicke model,'' in  a more restricted context,
partly because, as we are going to argue later, a  departure from their
assumption appears to be required eventually for the theory to be applied
successfully to the accelerating universe \cite{cmb}.

For the  later convenience let us give an illustrative example of $L_{\rm
 matter}$ for a free massive Dirac field as a convenient representative
 of matter fields;  
%%%%%%%%%%%%
\beq
 L_{\rm matter} =-\psibar \left( \partial\hspace{-.5em}/  +m\right)\psi,
 \quad\mbox{where}\quad m=\mbox{const}. 
\label{prs_7}
\eeq  
Due to the assumed absence of $\phi$, the mass of this Dirac field  is
 $m$, a pure constant.  Generally speaking, the  constancy of  masses of
 matter particles is a unique feature of the BD model.

By applying the conformal transformation $g_{\mu\nu} \rightarrow
g_{*\mu\nu}=\Omega^2 g_{\mu\nu}$ with a special choice $\Omega^2
=\xi\phi^2$, we move to what is called the  Einstein frame, in which
the same Lagrangian is re-expressed as 
%%%%%%%%%%
\beq
{\cal L}\hspace{-.3em}=\hspace{-.5em}\sqrt{-g_{*}}\left(\half R_{*} -
{\rm Sgn}(\zeta^{-2})\half g^{\mu\nu}_{*}\partial_{\mu}\sigma\partial_{\nu}\sigma - V(\sigma)  +L_{\rm *{\rm matter}}  \right),
\label{ipmu_15}
\eeq
without nonminimal coupling term, as designed, hence a purely constant
$G$ as in the standard Einstein-Hilbert term.

Note that the scalar field  $\phi$ in \reflef{ipmu_8}) and the same
$\sigma$ in \reflef{ipmu_15})  are related to each other  by
%%%%%%%%%%%%
\beq
\phi
=\xi^{-1/2}e^{\zeta\sigma},\quad\mbox{with}\quad \zeta^{-2}=6+\epsilon
\xi^{-1} =6+4\omega.
\label{ipmu_c3}
\eeq
Also the constant term $\Lmd$ in \reflef{ipmu_8}) has been converted to the
potential $V(\sigma) =\Lmd e^{-4\zeta \sigma}$ in \reflef{ipmu_15}).
Otherwise, we put the symbol $*$ nearly everywhere.   Also again,
however,  for the later convenience to discuss simple cosmology, we add 
%%%%%%%%%%%%%
\beq
a_* =\Omega a,\quad\mbox{and}\quad dt_*=\Omega dt,
\label{ipmu_c4}
\eeq
for the  scale factor $a$ and the cosmic  time $t$, respectively.
According to the first of \reflef{ipmu_c4}) the way of cosmological
expansion differs from frame to  frame.

We also transform the matter fields like \cite{ptp}
%%%%%%%%%%%%
\beq
 L_{\rm *matter} =-\psibar_* \left( \partial\hspace{-.5em}/
 +m_*\right)\psi_*,
 \quad \psi_*=\Omega^{-3/2}\psi, 
\label{prs_7a}
\eeq 
together with the transformation rule for the mass;
%%%%%%%%%%%
\beq
m_*=\Omega^{-1} m.
\label{ipmu_c5}
\eeq

We compare \reflef{ipmu_8}) with string theory.   For the closed
strings, we find an effective Lagrangian  
%%%%%%%%%%%%%
\beq
{\cal L}_{\rm string}= \sqrt{-\bar{g}}e^{-2\Phi}\left(
\half \bar{R} +2 g^{\bar{\mu}\bar{\nu}}\partial_{\bar{\mu}} \Phi
\partial_{\bar\nu} \Phi 
 -\frac{1}{12} H_{\bar{\mu}\bar{\nu}\bar{\lambda}}
H^{\bar{\mu}\bar{\nu}\bar{\lambda}}
\right),
\label{ipmu_18}
\eeq
as Eq. (3.4.58) of \cite{string}, in higher-dimensional spacetime.
Note the presence of a scalar field $\Phi$, called dilaton.   By
introducing $\phi =2e^{-\Phi}$, we can re-express the first two terms
with the result precisely the same as the first two terms in
\reflef{ipmu_8}),  provided $\epsilon =-1$ and $\xi =1/4$, or $\omega
=-1$.    This suggests that our Jordan frame corresponds to the world in
which unification is realized.  This also justifies the way of including
$\Lmd$ of the Plankian size in \reflef{ipmu_8}).  We may thus call the
Jordan frame a ``string'' frame or ``theoretical'' frame. Also it may
even appears as if Jordan's theory had to wait for decades before it was
rediscovered later by string theory.

Then how about the ``physical'' or ``observational'' frame?  According to
 Dicke  in this connection, the conformal transformation is a  local change of
 units \cite{dicke}.  Let us emphasize this view on the units.

Suppose we use an atomic clock,  measuring time in reference to the
frequency of certain atomic transition, in which we have the fundamental unit
provided by the electron mass $m_{\rm e}$.\footnote{The reduced mass
should be preferred in principle.  The required details are
straightforward, but will be avoided for the time being for simplicity. }  Then
we find no way to detect any change, if any, of  $m_{\rm e}$
itself, as long as we continue to use the atomic clock.  This might be
re-expressed by a more general term, ``own-unit-insensitivity
principle''\cite{ptp}.  Using atomic clocks then implies that  we are in
the physical frame in which $m_{\rm e}$ is kept constant.  According to
what we pointed out on the BD model, the constancy of $m_{\rm e}$
indicates the Jordan frame, implying the physical frame is precisely the
Jordan frame.  The same simple argument can be extended to a wider class
of astronomical observations based  on measuring redshift of atomic
spectra, again with the fundamental unit provided by $m_{\rm e}$ as in
using atomic clocks.  In this sense we repeat the above statement on the 
identification of the Jordan frame with the physical frame, as far as we
accept the BD model.

\section{Cosmology}

Let us go on to discuss cosmology in the presence of $\Lmd$ assumed
positive and of the Planckian size,  first in the Jordan frame.  As usual we 
make simplifying assumptions of the metric in the radiation-dominated
universe.  We  also assume $\phi$ to be spatially uniform depending only
on the cosmic time $t$.   We then write down equations and seek the
asymptotic solutions, given in Chapter 4.4.1 of \cite{cmb};  
\beqa
a&=&\mbox{const},\quad\mbox{or}\quad H=\frac{\dot{a}}{a} =0,
\label{jfr_1} \\
\phi &=& \sqrt{\frac{4\Lmd}{6\xi +\epsilon}}\: t,
\label{jfr_2}\\
\rho &=& -3\Lmd \frac{2\xi +\epsilon}{6\xi +\epsilon} = \mbox{const}.
 \label{jfr_3}
\eeqa
Skipping all the details, we focus upon \reflef{jfr_1}).  Rather
unexpectedly this implies a {\em static} universe.  We emphasize that
this solution  is not only asymptotic but
also an attractor solution which any solution starting with whatever
initial values tends to, as was reconfirmed by our recent reanalysis
\cite{kmyf}.  Due to this unrealistic aspect, the Jordan frame can be
hardly accepted as a physical frame, contrary to what we discussed before.

In this connection we point out that our solution fails to show a smooth
behavior in the limit $\Lmd \rightarrow 0$, suggesting that in the
presence of $\Lmd$, the solution can be different from what had been
derived in its absence, resulting in a drastic change.

Now in the Einstein frame the asymptotic solutions are obtained in
Chapter 4.4.2 of \cite{cmb};
%%%%%%%%%%%%%
\beqa
a_*&=&t_*^{1/2},\label{efr_1}\\
\sigma &=&\bar{\sigma}+\half \zeta^{-1}\ln t_*, \quad\mbox{with}\quad
\Lmd e^{-4\zeta \bar{\sigma}}=\frac{1}{16}\zeta^{-2},   \label{efr_2}\\
\rho_\sigma &=& \half\dot{\sigma}+V(\sigma) =
\frac{3}{16}\zeta^{-2}t_*^{-2},
 \label{efr_3}\\
\rho_*&=&\Omega^{-4}\rho=\frac{3}{4}\left( 1-\frac{1}{4}\zeta^{-2} \right) t_*^{-2}  \label{efr_4},
\eeqa
where $t_*$ is the cosmic time in the Einstein frame, obtained to be
$t_*\sim t^2$ in accordance with the second of \reflef{ipmu_c4}).  We
also have $\dot{\sigma} =d\sigma/dt_*$.

 According to \reflef{efr_1}) the universe now expands, as expected from
the discussion on the first of \reflef{ipmu_c4}),  precisely in the
same way as in the ordinary radiation-dominated universe,\footnote{This
turns out to be rather accidental, because the same behavior follows
even for the dust-dominance \cite{cmb,kmyf}.}  hence
tempting us to accept the Einstein frame as the physical frame.
However, examining \reflef{ipmu_c4}) and \reflef{ipmu_c5}) together
with the second of \reflef{prs_7}), we find\footnote{Combining these with
the first of \reflef{jfr_1}) leads to the condition $am =a_*m_*
={\mbox{const}}$, which implies that the universe, hence the inter-galactic
distances, grow with the same rate as the microscopic meter-stick
provided by $m^{-1}$ or $m_*^{-1}$, in totally inconsistency with
today's concept of the expanding universe. }
%%%%%%%%%%%%
\beq
m_*\sim t_*^{-1/2},
\label{efr_5}
\eeq
which fails to be constant obviously in contradiction with the
own-unit-insensitivity principle, a condition necessary to be a physical
frame.  We again face an unrealistic universe, no way to accommodate a 
physical frame.

\section{Leaving the Brans-Dicke model}

We then wonder if we find a way out by somehow  demanding a static $m_*$,
still keeping the expansion \reflef{efr_1})  as it is, so that we can accept
the Einstein frame as a physical frame.  This forces us finally to
decide to {\em leave} the BD model.  In fact we decided to revise our previous
choice of $L_{\rm matter}$ as in \reflef{prs_7}), by  replacing the
mass term by the Yukawa-type coupling, as shown by
%%%%%%%%%%%%
\beq
 L_{\rm matter} =-\psibar \left( \partial\hspace{-.5em}/  +f\phi\right)\psi,
\label{prs_77}
\eeq 
where $f$ is a dimensionless coupling constant.

In this way we derive a constant $m_*$, hence achieving the goal to
identify the Einstein frame as the physical frame.  The simple feature of
the  dimensionlessness of $f$ in \reflef{prs_77}) is shown to be shared
by any other terms in  the basic Lagrangian \reflef{ipmu_8}) except for 
the $\Lmd$ term.  In a sense we are having a global scale-invariance,
hence the name the ``scale-invariant'' model in place of the BD model
\cite{cmb}.

This is, however, not the end of the story.  We allowed $\phi$ to enter
$L_{\rm matter}$ in violation of the assumption due to Brans and Dicke.
For this reason we face the WEP violating terms, which fortunately turn
out to be unobservable in the classical limit.  But quantum 
effects arising from the interactions among matter fields will regenerate the WEP violating terms, though occurring somewhat suppressed
according to the estimates by means of {\em quantum anomalies}, a
well-established technique in the relativistic quantum field theory \cite{cmb}.
 We point out that this conclusion hinges upon the static universe solution
of the cosmological equation in the presence of $\Lmd$, as well as the
simple and straightforward analysis on how to define the physical
conformal frames.

Finally we come back to the Correlation, mentioned at the beginning.  As
we point out, an important clue is already found in \reflef{efr_3}) in the solution in the Einstein frame, which has been
shown to be a physical frame, where $\rho_\sigma$ is the dark energy
density, hence is interpreted as $\Lmd_{\rm eff}$, thus
giving\footnote{The same result had been obtained in \cite{yf82}.  See
also footnote 12 of \cite{yfnaha}, and the second paragraph of 
section 1 in \cite{kmyf}.}   
%%%%%%%%
\beq
\Lmd_{\rm eff}\sim t_*^{-2},
\label{ipmu_35} 
\eeq 
also to be called Scenario of a decaying cosmological constant.  This
Scenario does include the Correlation \reflef{ipmu_c2}) re-interpreted
in the Einstein frame, now  extended to much wider time
span.\footnote{Even wider than in \cite{freeze}.}  Obviously, we
reproduce our previous estimate $\Lmd_{\rm eff}\sim 10^{-120}$ for
$t_{*0}\sim 10^{60}$, though we no longer appeal  to
an extreme and unnatural fine-tuning process;  the left-hand side is
small nearly automatically, only because we are old  enough as indicated on the
right-hand side, somewhat {\em a la} Dirac \cite{dirac}.   Allowing us
to  deal with the number of this  small so naturally is certainly a
major  success of the scalar-tensor  theory, an advantage not shared by
many other phenomenological approaches.

Unfortunately this is still short of a complete success because the
smooth behavior $\rho_\sigma \sim t_*^{-2}$ in \reflef{efr_3})
is not the way we expect an extra acceleration of the universe, which we
are now watching.  But what we need can be only a relatively small
deviation off the dominant behavior, though in a rather phenomenological
way at this moment, probably with different laws for different kinds of
behaviors.

Without entering into any further details,
we show an example of the solutions plotted in Fig. 1, intended to be more
realistic \cite{cmb}.  In the bottom panel, $\rho_*$, the ordinary
matter density and  $\rho_s$, certain modification of $\rho_\sigma$ to
give $\Lmd_{\rm eff}$, are plotted against $\log t_*$.  Today is $\sim
60$.  We find two densities falling off like $t_*^{-2}$ as a {\em common
overall} behavior.  This is the  way we inherit the Scenario, but also
with sporadic interlacing behaviors, and ensuing extra accelerations of the
scale factor including the period around today, as exhibited in the top
panel.  They represent  {\em non-smooth} behaviors.   Summarizing we find that 
in spite of the  non-smooth aspects off the dominant overall behaviors,
the presence of  the underlying trend based on the simple scalar-tensor
theory is unmistakable.
\newpage     
 %%%%%%%%%%%
%\begin{figure}[h]
%\hspace{8.5em}
\bminip{7.8cm}
\hspace*{-1.6em}
\includegraphics[keepaspectratio,width=7.7cm]{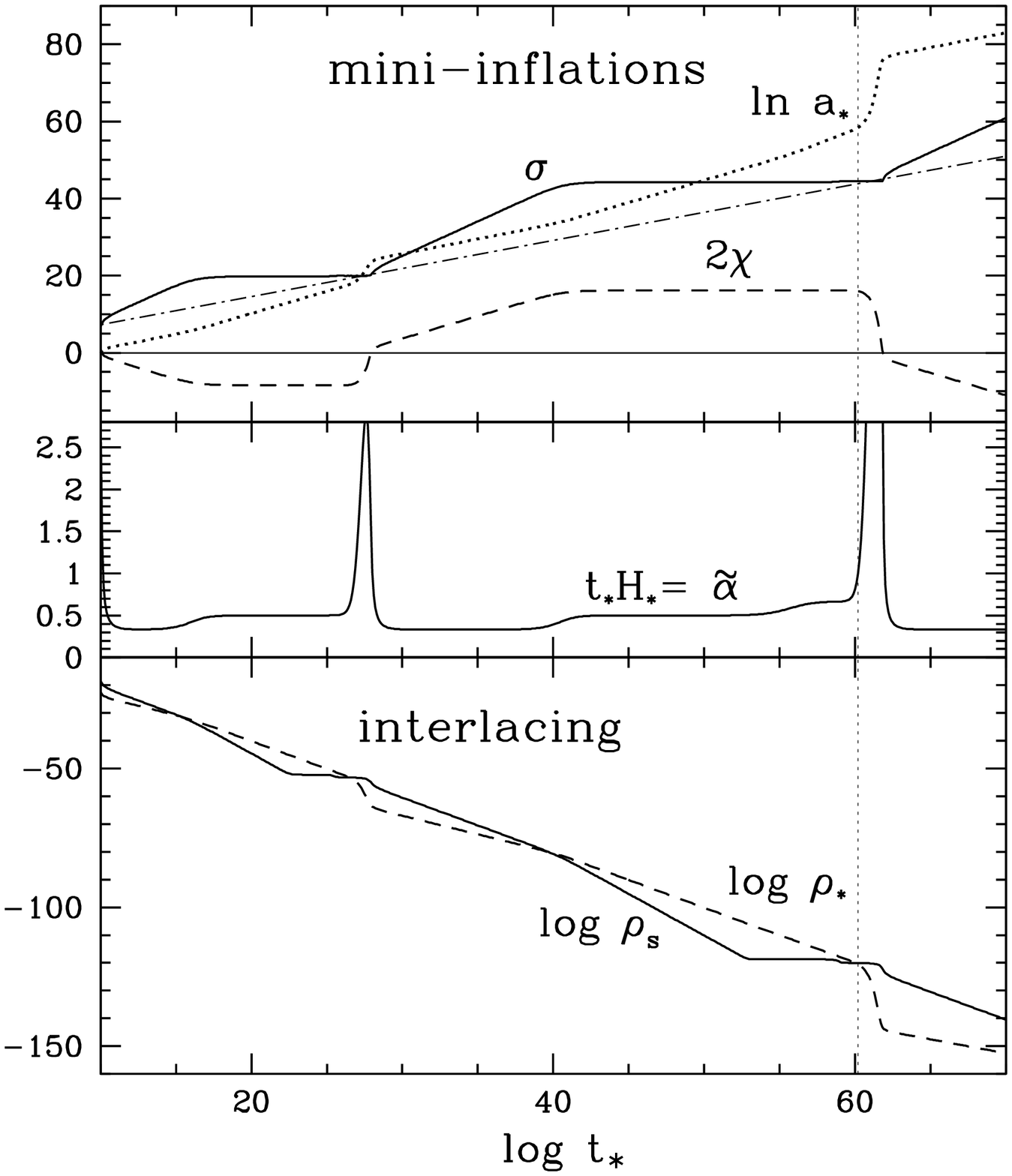}
\eminip
%\mbox{}\\[-1.4em]
\hspace{1.0em}
\bminip{6.7cm}
%\begin{figure}
Figure 1: Taken from Fig. 5.8 of [3].  In the bottom panel, we plot $\rho_*$,
the ordinary matter density in the Einstein frame, and  $\rho_s$, modified
 version of $\rho_\sigma$  including the contribution from the assumed
 added scalar field $\chi$, both in the logarithmic scale against $\log
 t_*$, which is $\sim 60$ today.  Two densities fall off with a common
 overall behavior $t_*^{-2}$, but with sporadic {\em interlacing} behaviors.
 Correspondingly we find sudden but finite increases of the scale factor
 $a_*$, called a {\em mini-inflations}, representing extra
 accelerations, as shown in the top panel, together also with
 $\tilde{\alpha}$  for the effective exponent of the scale factor shown
 in the middle panel.  
%\end{figure}
\eminip
\mbox{}\\[-.2em]


\begin{thebibliography}{99}
%\bibitem{}
\bibitem{accu}A.G. Riess {\em et al}. Astron. J. {\bf 116} (1998),
	1009. S. Perlmutter {\em et al}. Nat. {\bf 391} (1998), 51;
	Astrophys. J. {\bf 517} (1999), 565.
\bibitem{jordan}P. Jordan, {\it Schwerkraft und Weltall}, Friedrich
	Vieweg und Sohn, 1955.
\bibitem{cmb}Y. Fujii and K. Maeda, {\it The scalar-tensor theory of
	gravitation}, Cambridge University Press, 2003.
\bibitem{BD}C. Brans and R.H. Dicke, Phys. Rev. {\bf 124} (1961), 925.
\bibitem{ptp}Y. Fujii, Prog. Theor. Phys. {\bf 118} (2007), 983.
\bibitem{string}M.B. Green, J.H. Schwarz and E. Witten, {\it Superstring 
Theory}, Cambridge University Press, 1985.
\bibitem{dicke}R.H. Dicke,  Phys. Rev. {\bf 154} (1962), 2163.
\bibitem{kmyf}K. Maeda and Y. Fujii, Phys. Rev. {\bf D79} (2009), 084026.
\bibitem{yf82}Y. Fujii, Phys. Rev. {\bf D26} (1982), 2580.
\bibitem{yfnaha}Y. Fujii,  Proceedings of the Workshop on Cold
	Antimatter Plasmas and Application to Fundamental Physics,
	February 20-22, 2008, Naha, Okinawa,
	Japan. AIPConf. Proc. 1037:23-34, 2008; arXiv:0803.3103.
\bibitem{freeze}K. Freese {\em et al}. Nucl. Phys. {\bf B287} (1987), 797.
\bibitem{dirac}P.A.M. Dirac, Proc. Roy. Soc. {\bf A165} (1938), 199.
\end{thebibliography}
\end{document}